\documentstyle[aps,prl,floats,epsfig]{revtex}

\title{Observation of Supershell Structure in Alkali Metal Nanowires}
\author{A.I. Yanson$^{1}$, I.K. Yanson$^{1, 2}$, and J.M. van Ruitenbeek$^{1}$}
\address{$^{1}$Kamerlingh Onnes Laboratorium, Universiteit Leiden, \newline
Postbus 9504, NL-2300 RA Leiden, The Netherlands\newline $^{2}$B.
Verkin Institute for Low Temperature Physics and
Engineering,\newline National Academy of Sciences, 310164,
Kharkiv, Ukraine}

\begin{document}
\draft

\twocolumn[\hsize\textwidth\columnwidth\hsize\csname@twocolumnfalse\endcsname

\maketitle

\begin{abstract}
{Nanowires are formed by indenting and subsequently retracting two pieces of
sodium metal. Their cross-section gradually reduces upon retraction and the
diameters can be obtained from the conductance. In previous work we have
demonstrated that when one constructs a histogram of diameters from large
numbers of indentation-retraction cycles, such histograms show a periodic
pattern of stable nanowire diameters due to shell structure in the
conductance modes. Here, we report the observation of a modulation of this
periodic pattern, in agreement with predictions of a {\it supershell}
structure.}
\end{abstract}

\vskip2pc]

\narrowtext

\newcommand{\av}[1]{\mbox{$\langle #1 \rangle$}}

Metallic nanowires clearly exhibit quantum properties in their
conductance and structure. At the level of a few atoms in cross-section the
conductance through metallic nanowires can be described in terms of a finite
number of quantum modes, and it has been shown that this number is
determined by the number of valence orbitals of the metal atoms involved
\cite{scheer97,scheer98,cuevas98a}. For monovalent free-electron metals such
as gold and, in particular, the alkali metals the conductance in the
smallest contacts evolves roughly through the successive opening of distinct
conductance channels as the contact size increases \cite
{brom99,ludoph99,ludoph99b}. This quantum character of electrical transport
was already inferred from histograms of conductance values for atomic-size
point contacts, which show that the contacts have a preference for multiples
of the conductance quantum $G_{0}=2e^{2}/h$ \cite{brandbyge95,krans95},
after correction for a small series resistance.

It had been suggested that the formation of quantum modes in nanowires
should not only determine the conductance but also the cohesive energy \cite
{stafford97,ruitenbeek97,yannouleas97}. Recently, we have shown that the
stability of nanowires in the range of cross-sections up to about 130 atoms
is determined by electronic shell structure for quantum modes \cite{yanson99}%
. The shell structure was observed in conductance histograms, which were
measured by many times indenting and retracting two metal electrodes by
means of a mechanically controllable break junction (MCBJ) \cite{muller92}.
The key evidence for shell structure is a regular spacing of diameters for
wires with enhanced stability, where the diameters were obtained from the
semiclassical expression for the conductance \cite{torres94}. The shell
structure observed here is a close analogue of the shell structure observed
for alkali metal clusters \cite{brack93,deheer93}, and it is the same
principle that also applies to electrons in atoms and to protons and
neutrons in nuclei.

\begin{figure}[!b]
\begin{center}
\leavevmode
\epsfig{figure=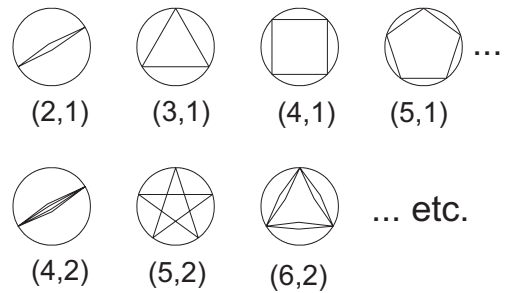,width=6.5cm}
\end{center}
\caption{Series of semiclassical orbits inscribed inside a circular
cross-section, which are applicable to spheres (clusters) and cylinders
(nanowires) alike. The orbits are labeled as $(M,Q)$, where $M$ is the
number of vertices and $Q$ is the winding number}
\label{fig:orbits}
\end{figure}

For metal clusters produced in vapor jets in vacuum, and analyzed by mass
selection, it was observed that clusters with certain `magic numbers' of
atoms, 8, 20 ,40 , 58, etc., are more abundant than others. This was
explained by their enhanced stability due to the closing of the shells of
electronic states, modeled as free electron waves confined to a spherical
potential well. The magic numbers can even be obtained from a semiclassical
expansion for the oscillating part of the density of single-particle levels \cite{brack93,balian72,brack97}, where the stable clusters are determined by
those diameters for which a bouncing electron wave traveling along a closed
classical path inscribed inside the spherical cluster walls obeys the
Bohr-Sommerfeld quantization condition with the bulk metal Fermi wave vector, $k_{F}$. The possible trajectories are illustrated in Fig.~\ref{fig:orbits}. It was shown that for spherical clusters the triangular and square orbits, with indices $(3,1)$ and $(4,1)$, dominantly determine the magic number series.

Since the two dominant orbits, triangle and square, lead to slightly different series of stable diameters, the interference between the two series gives rise to a beat pattern, known as {\it super\/}shell structure \cite{brack93,balian72,nishioka90}. For alkali metal clusters a single beat due to this effect has indeed been reported \cite{pedersen91,brechignac93}.
Here, we report the observation of supershell structure in sodium nanowires.
We show that, in contrast to clusters but in agreement with the theory for
nanowires, the diametric orbit (labeled $(2,1)$ in Fig.~\ref{fig:orbits})
has a strong contribution. Due to the larger separation between the periods
for diametric and higher orbits, several beat minima of the supershell
structure can be observed.

The experiment is performed using an MCBJ, modified to accommodate the very
reactive alkali metals \cite{krans95}. A rectangular piece of sodium metal
is cut, while immersed in paraffin oil for protection against oxidation, and
fixed upon a phosphor bronze bending beam with four 1\,mm screws. A notch is
cut in the center of the sodium sample, and the assembly is taken out of the
paraffin and quickly mounted in a sample holder in a three-point bending
configuration. Current and voltage leads are connected to both sides of the
notch and the sample holder is evacuated and cooled down to liquid helium
temperature. By applying force to the bending beam the sodium sample is
broken at the notch, by which two clean fracture surfaces are exposed.
Atomic-size contacts can then be established by relaxing the force, using a
piezoelectric element for fine control. A heater and thermometer permit
controlling the temperature from 4.2\,K to above 100\,K, while the vacuum
can remains immersed in liquid helium. The conductance of the contacts is
measured with a four-point dc-voltage bias circuit. The signal goes via a
current-to-voltage converter through a 16\,bit, $10^5$\,samples/s
analog-to-digital converter to a pc-based controller. The software also
drives the two halves of the sample into and out of contact by controlling
the piezovoltage. Conductance values are automatically accumulated into
histograms for typically over $10^4$ contact breaking cycles, with a
resolution of about 10 bins per $G_0$. In order to reduce digitization noise
a three bin wide smoothing function is applied to the histograms.

\begin{figure}[!t]
\begin{center}
\leavevmode \epsfig{figure=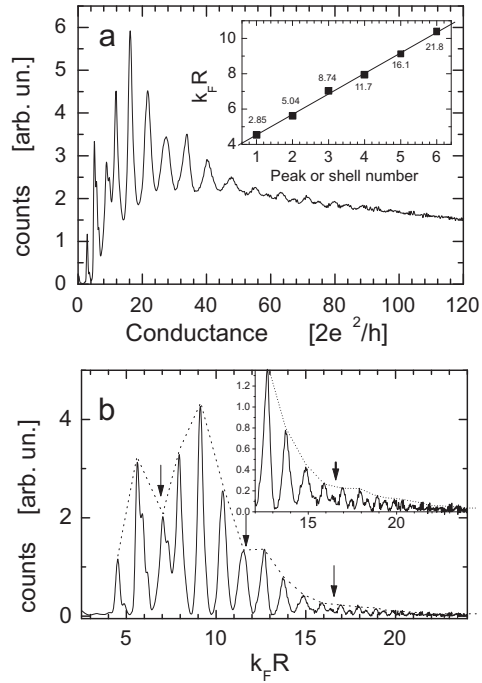,width=6.5cm}
\end{center}
\caption{Histogram of the number of times each conductance is
observed. (a) Data taken for sodium at $T=90$\,K and bias voltage
100\,mV, constructed from over 13800 individual indentation
cycles. The inset shows a plot of the radii of the first 6 maxima
versus the peak number, illustrating the regular spacing of the
peak positions (the numbers represent conductance values for each
peak in quantum units). (b) The conductance histogram for sodium
from (a) with a smooth background subtracted, plotted against the
radius. In the inset a magnified portion at high conductances is
displayed. The dotted curve envelops peaks in the histogram and
the arrows point to positions of minimal amplitude. These are the
nodes of the supershell structure. } \label{fig:histogram}
\end{figure}

At low temperatures and low voltage bias only four pronounced peaks at low
conductance are seen in the histogram, near 1, 3, 5 and 6\,$G_0$, which have
been attributed to the successive occupation of distinct quantum modes \cite
{krans95}. At higher conductance the histogram shows a number of rather wide
hills which grow into sharp peaks as we increase the temperature \cite
{yanson99}. Fig.~\ref{fig:histogram}a shows an example of a histogram for
sodium at a sample temperature of 90\,K. The histogram is similar to the one
presented in Ref.~\cite{yanson99}, but the modulation of the peak
intensities is more pronounced here \cite{note1}. The radius $R$ of the
nanowires can be obtained from the semiclassical expression for the
conductance \cite{torres94},
\begin{equation}
\frac G{G_0} =\left( \frac{k_FR}2\right) ^2\left( 1-\frac 2{%
k_FR}\right) .  \label{eq:torres}
\end{equation}
The inset in Fig.~\ref{fig:histogram}a shows the radii, in units $k_F^{-1}$,
corresponding to the first 6 maxima against their sequential number. Indeed,
for shell structure we expect regularly spaced peaks as a function of the
radius.

The modulation of the peak intensities and their periodicity is more clearly illustrated by subtracting a smooth background and plotting the histogram as a function of the radius, see Fig.~\ref{fig:histogram}b. Two beat minima are clearly visible at about $k_F R \simeq 7$ and 11.5, and a third can be seen on the expanded scale in the inset. Similar modulations have been observed in
histograms for potassium and lithium, which will be discussed elsewhere.

\begin{figure}[!b]
\begin{center}
\leavevmode
\epsfig{figure=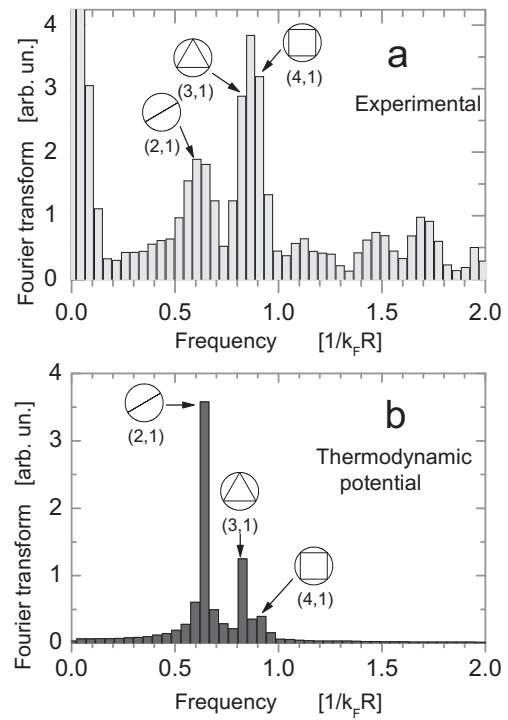,width=6.5cm}
\end{center}
\caption{(a) Fourier transform of the histogram of Fig.~\ref{fig:histogram}b
for sodium. Two main peaks at 0.6--0.65 and 0.8--0.9 demonstrate the
importance of the distinct contribution of the diametric orbit. (b) Fourier
transform of the calculated oscillatory part of the thermodynamic potential,
Eq.~\ref{eq:omega}, including terms $Q=1$ and $M=2,3,\ldots \,,7$. The shape
of the nanowire was set by the parameters $k_{F}L_{0}=200$ and $k_{F}A=50$.
The range of the variable $r$ is chosen to be approximately the same as for
the experimental histograms. The three main orbits, diametric, triangle and
square, are indicated.}
\label{fig:fourier}
\end{figure}

The components in the periodic structure are separated by making a Fourier
transform of the curve in Fig.~\ref{fig:histogram}b, which is shown in Fig.~%
\ref{fig:fourier}a. We observe two frequency components, one at 0.6--0.65
and the other at 0.8--0.9 $(k_F R)^{-1}$. We will argue that these are the
components of the supershell structure.

The quantum modes in a nanowire give rise to an oscillating contribution in the density of electronic states as a function of electron energy at constant radius or as a function of the radius of the wire at the Fermi energy. The modes each form a one-dimensional band, with quantum numbers determined by the confinement in the two transverse dimensions. This leads to an oscillating structure in the thermodynamic potential \cite{stafford97,ruitenbeek97,yannouleas97,yannouleas98,hoppler99,kassubek99}.
The latter determines the stability of a particular structure, which leads
to an oscillating probability for nanowire diameters. Also the Sharvin
conductance of the nanowires is modified by an oscillating contribution, but
this contribution appears to be of secondary importance to the experiment
\cite{yanson99}. The leading terms in a semiclassical expansion of the
oscillating part of thermodynamic potential are \cite{yannouleas98},
\begin{eqnarray}
& \Omega ^{osc}=\frac{2\varepsilon _{F}k_{F}}{\sqrt{2\pi r^{\prime \prime }}r}%
\sum_{M=2}^{\infty }\sum_{Q=1}^{M/2}\frac{1}{M^{7/2}}\left[ \sin
\left( \frac{\pi Q}{M}\right) \right] ^{-3/2} \times \nonumber \\
& \cos \left[ 2Mr\sin \left( \frac{\pi Q%
}{M}\right) +\frac{\pi }{2}\left( M+\frac{1}{2}\right) \right] ,
\label{eq:omega}
\end{eqnarray}
where $r=k_{F}R$, $r^{\prime \prime }=\partial ^{2}r/\partial z^{2}$ is the
curvature along the direction of the wire axis, and $r$ and $r^{\prime
\prime }$ are taken at the narrowest cross-section, $z=0$. The number $M$
corresponds to the number of vertices in the semiclassical orbit (see Fig~%
\ref{fig:orbits}) and $Q$ is the winding number. In order to evaluate this
expression we need to make some assumption about a smooth evolution of the
curvature of the wire, but this will only affect the prefactor and not the
oscillating structure itself. Following Ref.~\cite{yannouleas98} we assume a
parabolic wire shape, $r=r_{0}+4(A-r_{o})(z/L)^{2}$. This shape grows
smoothly from an initial cylindrical wire of length $L_{0}$ and radius $A$
while maintaining a constant volume, so that the radius at the narrowest
point can be expressed in terms of its length $L$ as $r_{0}=(A/4)(\sqrt{%
30L_{0}/L-5}-1)$. Substituting these expressions into Eq.~\ref{eq:omega} we
calculate the Fourier transform of $\Omega ^{osc}(r)$ in the same range of $r
$ as for the experimental data. The result is given in Fig.~\ref{fig:fourier}%
b, where we have included terms up to $M=7$ for $Q=1$ (the higher $Q$ give
rise to lower amplitude harmonics and we concentrate on the fundamental
components). The dominant peak is due to the diametric orbit $(M,Q)=(2,1)$
and the peaks for higher $M$ rapidly decrease in amplitude and their
frequency converges at 1, marking the end of the first series ($Q=1$). In
the Fourier transform we can only resolve two peaks in addition to the one
for $M=2$, corresponding to the triangular ($M=3$) and square orbits ($M=4$).

The experiment clearly shows the same groups of frequencies, one
corresponding to the diametric orbit and one peak at the position of the
triangular and square orbits, which cannot be individually resolved. Some of
the higher frequency maxima with small intensities are probably partly due
to harmonics of the main bands. The relative intensities of three calculated
peaks cannot be directly compared to the theoretical spectrum since the
latter results from the thermodynamic potential, which we do not measure
directly. The experimental spectrum is derived from the conductance
histogram and is expected to reflect the same components as the
thermodynamic potential, but the intensities depend, amongst others, on
kinetic factors for surface diffusion of atoms and the time available for
reaching the proper minima in the potential. Further differences may arise
from deviations in cylindrical symmetry of the wire and scattering of the
electrons on residual surface roughness. In particular the later mechanism
would favor the higher order orbits, since specular reflection increases for
smaller angle of incidence with the surface.

In contrast to cylindrical systems, for spherical systems the diametric
orbit ($M=2$) is negligible as it has a significantly smaller degeneracy
compared to the triangular ($M=3$) and squared ($M=4$) ones \cite{balian72}:
For a given position of one vertex the triangle and square have an
additional degree of freedom being the rotation around the normal to this
point. The contribution of the higher order orbits decays as a high power of
the index $M$ of the orbit. The closeness of the frequencies for the
triangular and square orbits leads to a long beating period. The first node
of the shell structure amplitude should be observed after approximately 12
periods of principal oscillations. The natural limit on the number of
chemical elements makes the number of shells in the periodic table too small
for supershell structure to be observable. The same applies to shell
structure in atomic nuclei. Although the amplitude of such a high number of
oscillations is greatly diminished, the first node of the beating pattern in
cluster abundance spectra was observed in Refs.~\cite
{pedersen91,brechignac93,pellarin93}.

For nanowires, the degeneracy of the diametric, triangle and square orbits
is of the same order and the larger separation between the $M=2$ frequency
and the higher ones make the beating pattern more readily visible in the
nanowire shell structure as compared to the metal cluster experiments. Thus,
for the first time a direct Fourier transform of the experimental data
yields proof for the existence of semiclassical orbits responsible for the
oscillations in the thermodynamic potential of the system as a function of
the radius.


\begin{references}
\bibitem{note1}  The details of the histograms are a.o. influenced by the
sample temperature and the bias voltage. Also, we expect that the crystal
structure of the banks at either side of the nanowire will be of importance.
We have no possibility to control this orientation, but it may be
responsible for the slight variation in appearance of the histograms. The
beat minima are not always equally pronounced, but their positions have been
reproduced for several samples, and the two main frequency components can
always be recognized in the Fourier spectrum.
\end{references}
\end{document}